\documentclass[twocolumn,secnumarabic,amssymb, nobibnotes, aps, prl, superscriptaddress]{revtex4-2}

\usepackage{graphicx}% Include figure files
\usepackage{dcolumn}% Align table columns on decimal point
\usepackage{bm}% bold math
\usepackage{amsmath}
\usepackage{enumerate}
\usepackage{setspace}
\usepackage{dsfont}
\usepackage{subfigure}
\usepackage{multirow}
\usepackage{indentfirst} %段落缩进值宏包
\usepackage {mathrsfs}%花体字
\usepackage{color}%字体颜色

\usepackage{lineno}

\usepackage{makecell}   %表格
\usepackage{multirow}	%表格

\usepackage{threeparttable}  %表格注释
\usepackage[pdfstartview=FitH,
CJKbookmarks=true,
bookmarksnumbered=true,
bookmarksopen=true,
colorlinks, 
pdfborder=001,
linkcolor=blue,
anchorcolor=blue,
citecolor=blue
]{hyperref}

\begin{document}
	
%\linenumbers

\title{Benchmark for Non-reciprocal Attack Detection on Synchronization}%

\author{Zishuo Ren}%
\affiliation{State Key Laboratory of Photonics and Communications, School of Electronics, and Center for Quantum Information Technology, Peking University, Beijing 100871, China}

\author{Yiting Lin}%
\affiliation{State Key Laboratory of Photonics and Communications, School of Electronics, and Center for Quantum Information Technology, Peking University, Beijing 100871, China}

\author{Yufei Zhang}%
\affiliation{State Key Laboratory of Photonics and Communications, School of Electronics, and Center for Quantum Information Technology, Peking University, Beijing 100871, China}

\author{Yang Li}%
\affiliation{Science and Technology on Communication Security Laboratory, Institute of Southwestern
Communication, Chengdu 610041, China}

\author{Ziyang Chen}%
\email[E-mail: ]{chenziyang@pku.edu.cn}
\affiliation{State Key Laboratory of Photonics and Communications, School of Electronics, and Center for Quantum Information Technology, Peking University, Beijing 100871, China}

\author{Hong Guo}%
\email[E-mail: ]{hongguo@pku.edu.cn}
\affiliation{State Key Laboratory of Photonics and Communications, School of Electronics, and Center for Quantum Information Technology, Peking University, Beijing 100871, China}

%\date{September 2022}%
%\date{\today}%

\begin{abstract} 
%modified
A precise and secure time synchronization is the backbone of both fundamental physics and advanced technologies. Despite ultra-high precision, security, particularly the unresolved vulnerabilities on physical links beyond traditional cryptography, remains the bottleneck. Here, we develop an analysis framework, the Benchmark Attack Noise Detection (BAND) model, to quantify the security of the high-precision time synchronization when its core assumption, reciprocity, is broken. The model characterizes non-reciprocal attacks through a unified metric, termed attack intensity. Based on the analysis of attack intensity, we accurately predict both instantaneous and accumulating attacks in a 100-km dual-comb two-way time-transfer system. Beyond security considerations, we find that the BAND model can explain intrinsic fiber noise and delay-unsuppressed noise naturally, offering an effective tool to establish an environment-sensitive link model. This work paves the way for a secure large-scale integrated time, sensing and communication networks.
%before modification
%A precise and secure time synchronization is the backbone of both fundamental physics and advanced technologies. Despite ultra-high precision, security, particularly the unresolved vulnerabilities on physical links beyond traditional cryptography, remains the bottleneck. Here, we develop an analysis framework, the Benchmark Attack Noise Detection (BAND) model, to quantify the security of the high-precision time synchronization when its core assumption, reciprocity, is broken. The model characterizes non-reciprocal attacks through a unified metric, termed attack intensity. Based on the analysis of attack intensity, we accurately predict both instantaneous and accumulating attacks in a 100-km dual-comb two-way time-transfer system. Beyond security considerations, we find that the BAND model can explain intrinsic fiber noise and delay-unsuppressed noise naturally, offering an effective tool to establish an environment-sensitive link model. This work paves the way for a secure large-scale integrated time, sensing and communication networks.

\end{abstract}

\maketitle
%\tableofcontents  %目录

%\section{INTRODUCTION}\label{sec1}

\begin{figure*}%[t]
	\centering
	\includegraphics[width=0.75 \linewidth]{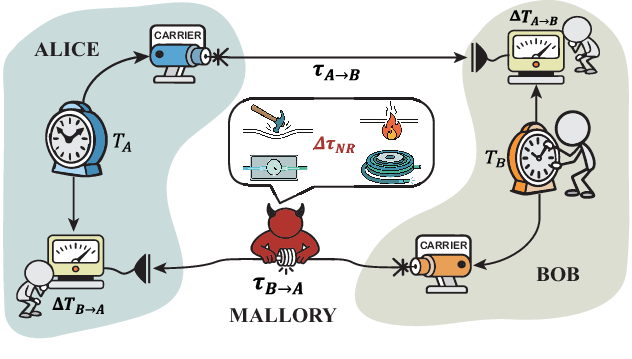}
	\caption{High-precision time transfer between non-portable clocks relies on the Einstein synchronization protocol, which exchanges time information through periodical pulses of the same frequency between spatially separated clocks Alice and Bob. This protocol is founded on the assumption that the propagation delay within the synchronization channel is symmetric in both directions. However, this assumption is not always valid in real-world scenarios, as potential adversaries Mallory can deliberately introduce non-reciprocal time delays $\Delta\tau_{\text{NR}}$ to deceive Alice and Bob into accepting incorrect timing information $\left(\Delta T_{A\rightarrow B}-\Delta T_{B\rightarrow A}+\Delta\tau_{\text{NR}}\right)/2$, rather than $\left(\Delta T_{A\rightarrow B}-\Delta T_{B\rightarrow A}\right)/2$, without any awareness of the discrepancy.}\label{Einstein protocol}
    \end{figure*}

{\textit{Introduction}}---With the rapid development of information society, the demand for a secure precise time is ubiquitous. Far beyond its role as the bedrock of positioning, navigation, timing (PNT) systems~\cite{JournalofGeodesy_2015,Lewandowski_2011,JournalofGeodesy_2009} and network security~\cite{110421}, it also constitutes both fundamental and frontier physical research~\cite{caldwell2023quantum,gozzard2022ultrastable,collaboration2021frequency,RevModPhys.90.045005,delva2017test,lisdat2016clock,Nat.Phys.10.933.2014}. On top of that, it enables the development of next-generation quantum networks~\cite{komar2014quantum,kimble2008quantum}, which integrate multiple functionalities such as quantum key distribution~\cite{luo2023recent}, distributed quantum sensing~\cite{guo2020distributed}, and quantum computing~\cite{main2025distributed}. All these applications inherently rely on two critical aspects of time–frequency technology: performance and security. The state-of-the-art time synchronization has achieved femtosecond-level precision~\cite{chen2025time,chen2024dual,yu2023time,shen2022free,deschenes2016synchronization,droste2013optical,giorgetta2013optical,lee2010time}, which is sufficient to support the aforementioned applications. However, as is shown in Fig.~\ref{Einstein protocol}, malicious interference that induces unpredictable or undetectable timing errors in the synchronization protocol~\cite{einstein1905electrodynamics} will nullify the benefits of the high-performance time applications.

%that may compromise or disable the aforementioned technologies.

    Previous studies identified non-reciprocal attacks, which fundamentally violate the bidirectional reciprocal assumption of two-way synchronization protocols, representing the most formidable threats to defend against~\cite{narula2018requirements}. Such attacks on physical signals, different from attacks on data information~\cite{dai2020towards}, render conventional cryptographic or data-layer protections ineffective. They have demonstrated severe damage in widely used network time synchronization~\cite{mizrahi2012game,mills2010network,ullmann2009delay} and satellite-based synchronization systems~\cite{picciariello2024quantum,jafarnia2012gps}. For next-generation synchronization based on optical methods~\cite{riehle2017optical}, although precision is greatly improved, non-reciprocal attacks are easier to execute and can cause even more significant damage~\cite{lee2019asymmetric}. In recent years, several efforts~\cite{xiang2025towards,li2023secure} have been proposed, trying to mitigate non-reciprocal attacks in a specific form. Unfortunately, to the best of our knowledge, existing reported research, in both classical~\cite{chen2024dual,he2018long,kim2008drift} and quantum schemes~\cite{liu2021quantum,quan2016demonstration,valencia2004distant,giovannetti2001clock}, does not provide a comprehensive analytical framework for non-reciprocal attacks. Owing to the absence of effective theoretical framework, the attacks remain inherently arbitrary and elusive, and thus are extremely difficult to detect or defend against. The fundamental challenge lies in the lack of appropriate physical constraints, which makes it difficult to quantitatively characterize and classify attack behaviors.
    
    In this Letter, we propose the Benchmark Attack Noise Detection (BAND) model for identifying non-reciprocal attacks. Remarkably, by imposing physical constraints in practical experiments, the attack behavior converges toward a Markovian structure in the time domain. Based on this, we show that the attacks manifest solely as variations in two types of noise: white phase noise and white frequency noise, and the BAND model quantifies the resulting system degradation through a unified metric termed “attack intensity”. Within this framework, attacks are identified by contrasting the post-attack against the attack-free time deviation (TDEV), which serves as the baseline describing the secure state of the system. We successfully detected various non-reciprocal attacks in a 100-km dual-comb two-way optical  
time synchronization system,  including instantaneous disturbance attack, and accumulating delay attack, with discrepancies between theoretical and measured attack intensities remaining within 5\%. Moreover, the BAND model can be extended to give a quantitative interpretation of fiber intrinsic noise~\cite{williams2008high,newbury2007coherent} and delay-unsuppressed noise~\cite{yu2024microwave,schioppo2022comparing,droste2013optical,williams2008high,newbury2007coherent}, constituting an essential advancement in optical fiber link modeling.

%\section{Benchmark of ADA Noise Detection (BAND) model}\label{sec2}
%\subsection{Postulates of BAND model}
\textit{Benchmark of attack noise detection model}---To analyze the characteristic of the non-reciprocal attacks, we introduce a model called benchmark attack noise detection (BAND). To formally analyze the detectability and statistical behavior of attacks under practical constraints, we introduce three foundational postulates that specify the minimal and realistic assumptions on measurement resolution, attacker behavior, and protocol compliance:

\begin{enumerate}
    \item Postulate 1. \textbf{Limited Resolution of Delay Measurement.} Due to the limited measurement resolution at the user end, the link delay variation can only be detected when it is greater than the resolution $\mathcal{T}_0$.
    \item Postulate 2. \textbf{Markovian Property of Delay Attack.} The user cannot know the specific form of the attack in advance and cannot infer future attacks based on past attacks.
    \item Postulate 3. \textbf{Cautiousness of Delay Attack.} In order to carry out an attack while concealing their intentions as much as possible, the attack behavior must first comply with the Einstein Protocol~\cite{einstein1905electrodynamics}, which means the frequency of the attack should be stationary. Otherwise the obvious change in signal exchanging rate is quite obvious and can be directly detected.
\end{enumerate}
   
Based on the above assumptions, BAND model is divided into two dimensions. One is to model of the sampling time dimension ${t}$. The other is to model the measured clock offset dimension ${\Delta\tau}$.
%\subsection{Effective attack points (EAP)}

\textit{Effective attack points}---In the sampling time dimension, according to postulate 1, non-reciprocal attack $\Delta\tau_{\text{attack}}(t)=\Delta \tau_{\text{NR}}(t)/2$ can only be observed by the user at discrete moments (see Supplemental Material~\cite{SM}). We name these moments {effective attack points (EAPs)}, denoted by $\left\{t_i \mid i \in \mathbb{N}^{+}\right\}$. According to postulate 2, the intervals $\left\{\delta t_i=t_{i+1}-t_i \mid i \in \mathbb{N}^{+}\right\}$ between any two adjacent EAPs are statistically independent random variables. According to postulate 3, the EAPs should be uniform in time sequence, without obvious dense or sparse distribution. That means the expectation value of $\left\{\delta t_i \mid i \in \mathbb{N}^{+}\right\}$ does not change over time: say $E\left[\delta t_i\right]=t_0$. According to maximum entropy principle, $\delta t_i \sim \exp \left({1}/{t_0}\right)=\exp \left(\lambda_0\right)$, where $\lambda_0$ [Hz] represents the number of
EAPs per unit time. As a consequence, the total number of EAPs per unit time $\Delta t$ should follow a Poisson distribution: $N(\Delta t) \sim \operatorname{Poisson}\left(\lambda_0 \Delta t\right)$~\cite{SM}. Therefore, the non-reciprocal attack $\Delta\tau_{\text{attack}}(t)$ can be modeled at the user end using the homogeneous Poisson process.

Notably, there may also be certain specific attack strategies whose EAPs have temporal aggregation and sparsity while the signal exchanging rate approximately remains stationary. In these cases, the expectation value of $\{\delta t_i\}$ is no longer a constant over time. For these attacks, we can model them using non-homogeneous Poisson processes. For general considerations, this article mainly discusses the homogeneous ones, and we leave the non-homogeneous cases for further discussion.

%\subsection{Attack kernels}
\textit{Attack kernels}---In the measured clock offset dimension, we introduce the attack kernel $K(t,t_i)$ to describe the impact that each EAP of
$\Delta\tau_{\text{attack}}(t)$ casts on the offset measurement. According to whether this impact accumulates in the subsequent measured clock offset, there exists two types of attack kernel: 

    \textbf{Type I:} The attack only affects the measurement of the clock offset in a very short period of subsequent time. 
    
    \textbf{Type II:} The impact of the attack accumulates in all of the subsequent clock offset measurements.

According to postulate 1, 
%the measured clock offset values at an EAP satisfies：$\left|\Delta \tau_{\text {attack }}\left(t_i\right)\right|=\mathcal{T}_0$. 
the type I kernel can be written as $K_1(t,t_i)=\pm\mathcal{T}_0\delta(t-t_i)$, and the type II kernel can be written as $K_2(t,t_i)=\pm\mathcal{T}_0U(t-t_i)$, where $\delta(t-t_i)$ represents delta function and $U(t-t_i)$ represents the unit step function.
%\subsection{Attack behavior}

\textit{Attack behavior}---Combining the EAP and attack kernel, the model for non-reciprocal attacks can be described by
\begin{equation}
\Delta \tau_{\text {attack }}(t)=\sum_{i=1}^{N(t)} K_{r_i}\left(t, t_i\right), r_i=1 \text { or } 2
\end{equation}
where $N(t) \sim \operatorname{Poisson}(\lambda t)$.

This model allows for simulation of different asymmetric delay attacks. For example, one can simulate four kinds of basic attack behaviors---phase jumping or jittering by selecting $r_i=1$, and simulate phase tampering or rambling by selecting $r_i=2$. Arbitrary attacks can be obtained through the superposition of these four types of attack based on the different choices of $\{r_i\}$ sequence.

\textit{Noise feature and attack intensity}---Now let us discuss the noise introduced by the basic attacks. We can obtain that the power spectral density (PSD) of the clock offset measurement caused by $\Delta\tau_{\text{attack}}(t)$ when $\{r_i \}=1$ satisfies $ S_{\text{attack}} (f) \propto \lambda\mathcal{T}_0^2$~\cite{SM}. Thus the noise introduced by the phase jitter or phase jumping on the measurement is white phase noise. Meanwhile, the attack induced PSD of the clock offset measurement when $\{r_i \}=2$ satisfies $ S_{\text{attack}} (f) \propto \lambda\mathcal{T}_0^2\cdot f^{-2}$. Thus the noise introduced by the phase tampering or rambling in the measured phase time is white frequency noise.

According to their noise feature, we can define a unified metric called attack intensity, to describe the influence of non-reciprocal attacks on a time synchronization system. Expressed as $I=\lambda\mathcal{T}_0^2$[s], it has a dimension of time, and can be directly obtained through the PSD. In real synchronization systems, due to the limited sampling rate, $\Delta \tau_{\text {attack}}$ is discretized by factor $\mathcal{T}\geq \mathcal{T}_0$. Therefore, the theoretical attack intensity should be $I_{\text{theo}}=\lambda\mathbb{E}\left\{\mathcal{T}^2\right\}$~\cite{SM}. $I_{\text{theo}}$ represents the ``power" of the delay variation introduced by the attack. As we will see, it is also a useful tool for the identification of non-reciprocal attack.

\textit{Attack identification}---Let’s discuss the identification of attacks in real clock synchronization systems. In practical experiments, TDEV is one of the most widely used performance metrics in the time-frequency fields for precision and stability assessment. More importantly, by smoothing the PSD, it enables a more accurate discrimination among different types of power-law noise. Therefore, we use TDEV to probe the noise changes caused by the attacks.

Based on several kinds of power-law noise, the TDEV of a time-frequency synchronization system can be decomposed as:
%\begin{equation}
%\begin{aligned}
%&\operatorname{TDEV}\left(\tau\right)=\\
%&\sqrt{\left(C_{\mathrm{wp}} \cdot \tau^{-\frac{1}{2}}\right)^2+C_{\mathrm{fp}}^2+\left(C_{\mathrm{wf}} \cdot \tau^{\frac{1}{2}}\right)^2+\left(C_{\mathrm{ff}} \cdot \tau\right)^2+\left(C_{\mathrm{rw}} \cdot \tau^{\frac{3}{2}}\right)^2}
%\end{aligned}
%\end{equation}

\begin{equation}
\operatorname{TDEV}(\tau)
=
\sqrt{
\sum_{i=-4}^{0}
\left( C_i \, \tau^{\frac{-(i+1)}{2}} \right)^2
},
\end{equation}
where $i=~ 0,~-1,~-2,~-3,~-4$ represent white phase noise, flicker phase noise (pink noise), white frequency noise (Brownian noise), flicker frequency noise and random walk frequency noise respectively. $C_i$ represents the contribution of the corresponding noise to TDEV. The relationship between these coefficients and the corresponding PSD of noises is given in~\cite{SM}.

According to the non-reciprocal attack model, attack will lead to an increase in white phase noise $(i=0)$ and white frequency noise $(i=-2)$ only. Therefore, the intensity changes of the two types of noise can be used to determine whether the system is secure or not. By using the particle swarm optimization algorithm, we calibrate a set of intensity coefficients $C_{0}^{\text{secure}}$ and $C_{-2}^{\text{secure}}$ of white phase noise and white frequency noise during the attack-free operation of the system as the background noise. For the specific coefficient calibration methods, see~\cite{SM}. When the system is under attack, the noise intensity coefficients become $C_{0}^{\text{attack}}$ and $C_{-2}^{\text{attack}}$. The attack intensity can thus be obtained by taking the difference of the noise coefficients before and after the attack.

To demonstrate the validity of this model, we conducted experiments of several common types of asymmetric delay attack using the dual-comb two-way optical time synchronization system~\cite{chen2024dual} in our laboratory.

\textit{Instantaneous disturbance attack}---The characteristics of instantaneous disturbances are quick recovery and short duration in experiments. Therefore, this type of attack can be described by the type I attack kernel.

\begin{figure}[h]
	\centering
	\includegraphics[width= 0.9\linewidth]{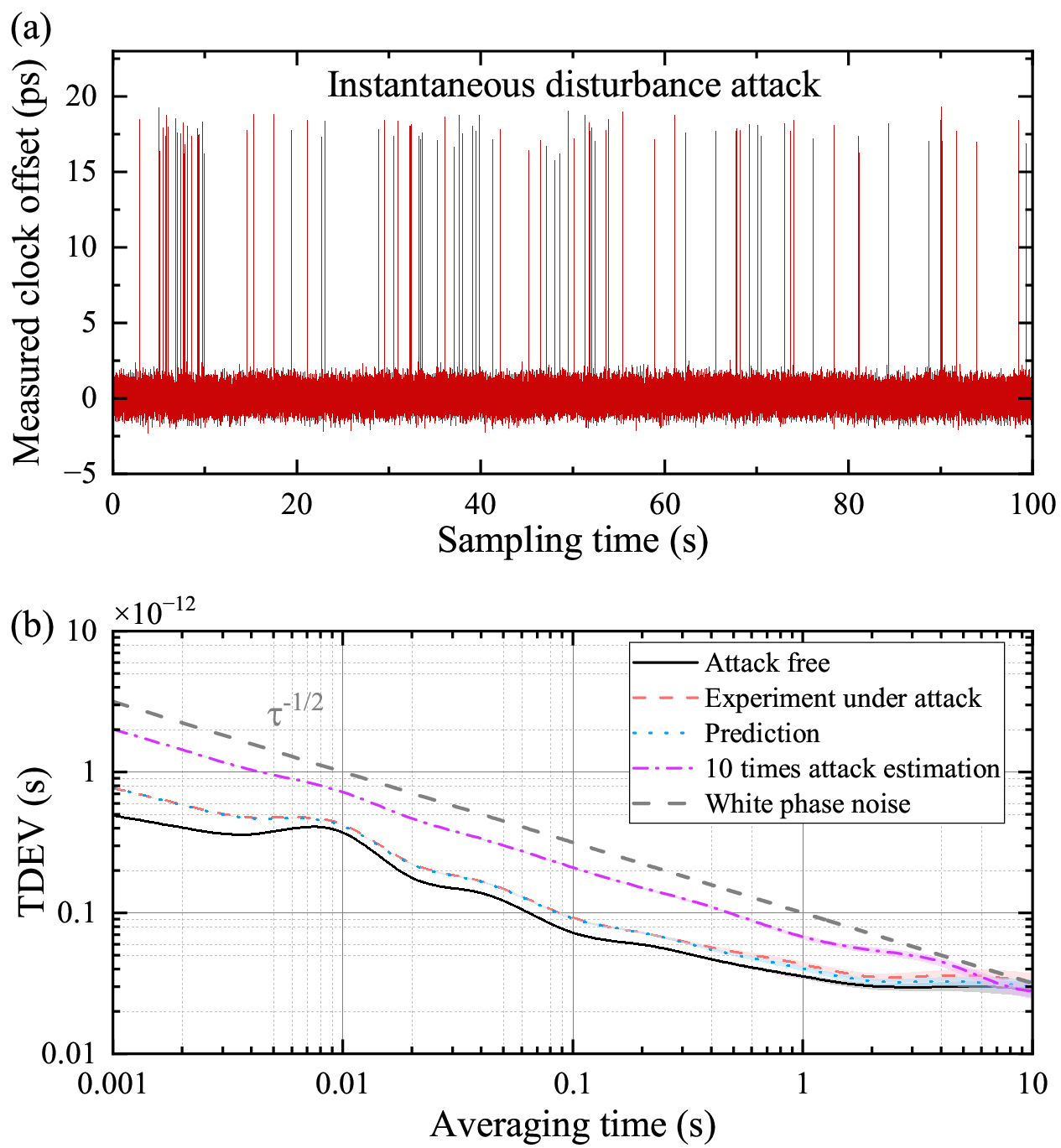}
	\caption{Instantaneous disturbance attack. (a) Attack behavior. (b) TDEV comparison: secure system (black), attacked system (orange), model prediction (blue), and estimated $10I_{\text{theo}}$ (purple).}\label{k1}
\end{figure}

%The calculation method of its theoretical attack intensity is shown in Appendix F.
We first carried out an instantaneous disturbance attack by tapping the connection joints of 100 kilometers of spooled optical fiber, and its time-domain behavior is shown in Fig.~\ref{k1} (a), and the red line is the attacked TDEV. We obtained $C_{0}^{\text{attack}}$ as $1.09\times 10^{-27}$ s. By comparing with the attack-free system whose noise intensity $C_{0}^{\text{secure}}$ is obtained as $6.96\times 10^{-28}$ s, we obtained an attack intensity $I_{\text{exp}}=3.93\times10^{-28}$ s.

On the other hand, we found that the characteristics of this attack are quick recovery and short duration. Therefore, this type of attack should be described by the type I attack kernel. By calculating $\lambda\mathbb{E}\left\{\mathcal{T}^2\right\}$, we obtained $I_{theo}=3.75\times10^{-28}$ s~\cite{SM}, with a deviation of $4.8\%$ compared to $I_{\text{exp}}$. Proved the instantaneous disturbance attack can be successfully predicted by our type I kernel. To show the greater impact on system, we simulated an attack with $10$ times of $I_{\text{theo}}$, as is shown in the blue line. Indicating a large boost in white phase noise and thus a great destruction in the short-term stability.

\begin{figure}[h]
	\centering
	\includegraphics[width= 0.9\linewidth]{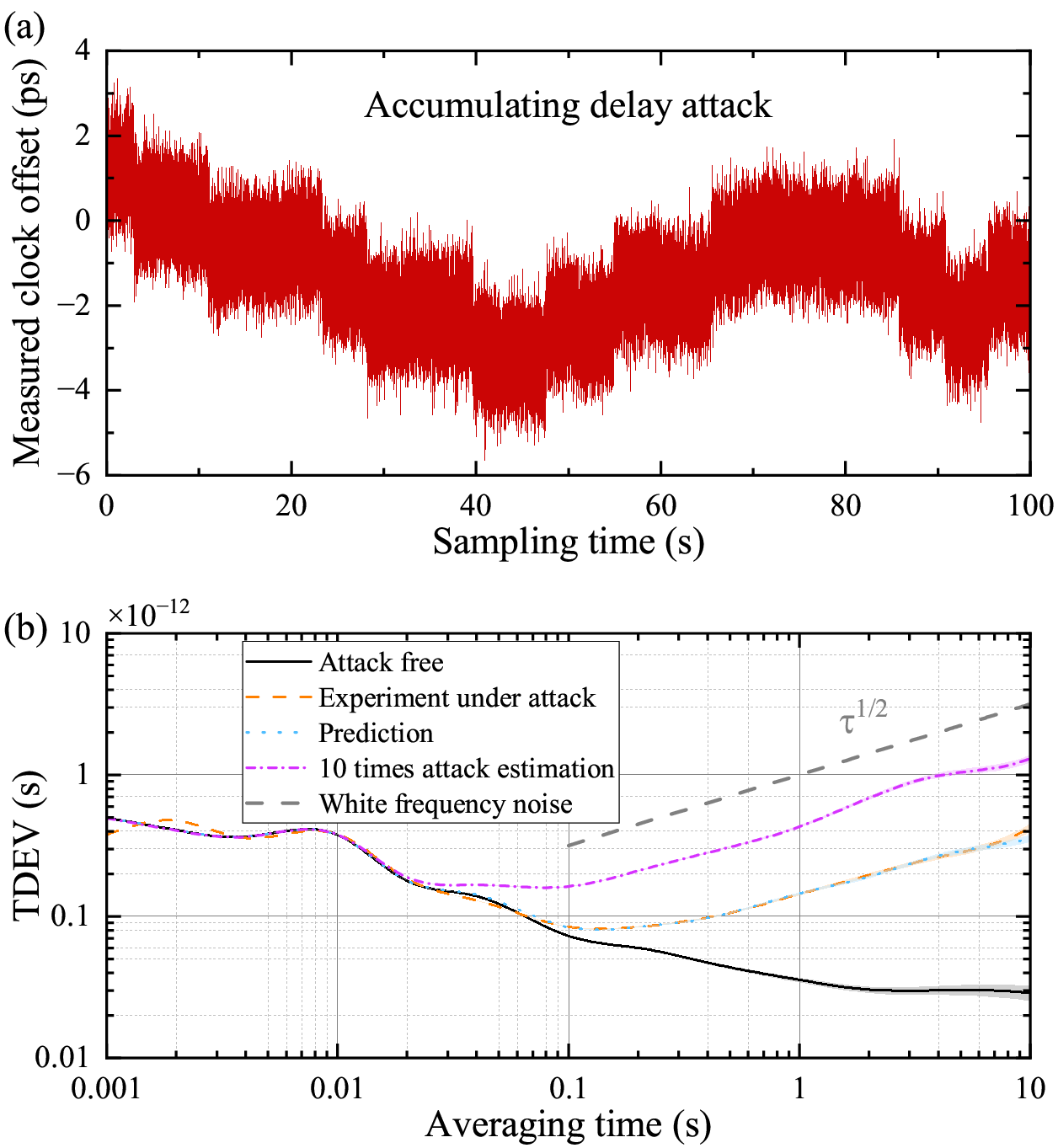}
	\caption{Accumulating delay attack. (a) Attack behavior. (b) TDEV comparison: secure system (black), attacked system (orange), model prediction (blue), and estimated $10I_{\text{theo}}$ (purple).
	}\label{k2}
\end{figure}

\textit{Accumulating delay attack}---We then carried out the accumulating delay attack by adding a small section of controllable optical delay line. Its time-domain behavior is shown in Fig.~\ref{k2} (a). We obtained  $I_{\text{exp}}=1.21\times 10^{-25}$ s by differencing $C_{-2}^{\text{secure}}$ and $C_{-2}^{\text{attack}}$. The characteristic of this attack is accumulative effect on the system, thus it should be modeled by type II kernel. We obtained $I_{\text{theo}}=1.16\times 10^{-25}$ s with a deviation of $4.8\%$~\cite{SM}. By simulation, we found the larger $I_{\text{theo}}$ of type II kernel leads to the greater destruction in long-term stability. 

In addition, by changing $\lambda$ or $\mathcal{T}$, we found that the influence on the stability of the system is unchanged as long as $I_{\text{theo}}$ is kept invariant. This shows the validity of $I$ in measuring the impact of attack on system performance, and further indicates that, in certain cases, single-point detection methods~\cite{li2023secure} may fail, making it necessary to employ the statistical detection techniques given by the BAND model.

\textit{The extension of BAND model}---We further verified the validity of BAND model beyond the attack scenario, as in studying the noise feature of a long distance frequency synchronization transfer in fiber. The noise impact of long transmission distances on the system has two aspects, the temperature-induced noise and the inevitable noise due to the travel time of transmitted signals. These two aspects could be treated as equivalent non-reciprocal attacks~\cite{SM} and could be analyzed using BAND model respectively. We used a 3000-km comb-based frequency transfer system~\cite{yu2024microwave} to conduct our experiments for clarity of the results. 

\begin{figure}
	\centering
	\includegraphics[width= 1\linewidth]{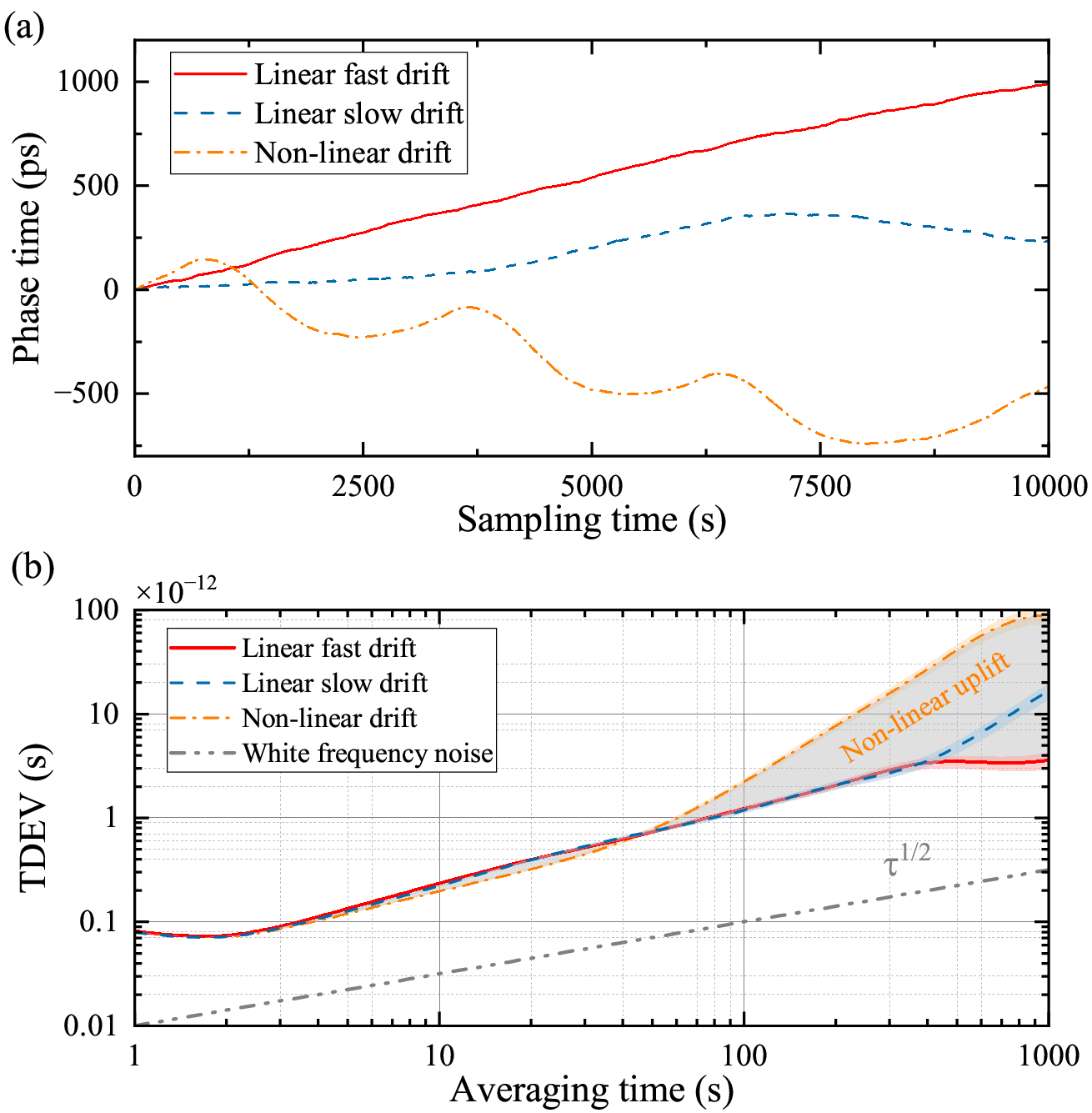}
	\caption{One-way signal under different temperature conditions. (a) Phase time domain behavior. (b) TDEV of linear fast drift (red), linear slow drift (blue) and non-linear drift (yellow).
	}\label{temp}
\end{figure}

\textbf{Temperature-induced noise}. First, the temperature-induced effect can be observed directly from the one-way transfer signal without any compensation. Experimentally, we measured several sets of phase data with the free-running system under different temperature conditions. The time domain phase data and their TDEV are shown in Fig.~\ref{temp}. The TDEV from 1 s to 100 s are almost the same regardless of the change of temperature. This white frequency noise corresponds to the so-called intrinsic fiber noise (IFN) observed in many previous works~\cite{williams2008high,newbury2007coherent}. After 100 s, noise introduced by 2nd-ordered and above non-stationary drift overtakes IFN which is generated by stationary drift, as is shown by the orange curve.

Theoretically, this one-way behavior can be equivalently interpreted as a two-way mechanism, in which one direction incorporates an additional temperature asymmetry $\Delta T_{\text{NR}}$ produced by 3000-km fiber. Since the non-reciprocal influence, $\Delta \tau_{\text{NR}_T}$, is accumulative, this process can be described by the type II attack kernel. The theoretical attack intensity in this scenario is given by $I_{\text{theo}}=\tau_0\cdot\mathbb{E}\{y^2\}$, and is is consistent with the experimental attack intensity~\cite{SM} with a deviation below $10\%$ for linear drift. This deviation grows as the linearity of the drift declines. In the extreme case where the non-stationary components fail postulate 3, the corresponding noise will leads to a deviation of $73\%$. It goes down to $9\%$ after considering the non-homogeneous correction. 

The IFN is produced by the temperature drift since the change of phase time signal is linear to the temperature change of fiber (see~\cite{SM})~\cite{yu2024microwave,sliwczynski2010optical}. Based on BAND model, we can provide a physical explanation for IFN: the thermal motion of a single particle in circumstance can be regarded as an instantaneous disturbance to the optical fiber, and the impact it contributes to phase time is described by type I kernel. The thermal motion of a large number of independently moving particles will integrate this disturbance, and therefore produces an ``attack" described by type II kernel, whose amplitude is proportional to temperature change. The IFN we observed is the 1st-order stationary part of this attack. We can further derive the relation between the PSD of IFN and temperature of fiber as 
\begin{equation}
    S_{\text {fiber}}(f)=2 \tau_0\left(\frac{f_0}{f}\right)^2 \alpha^2 L^2 \mathbb{E}\left\{\left(\frac{\Delta T}{\Delta t}\right)^2\right\},
\end{equation}
where $\alpha$ is the temperature coefficient of fiber~\cite{SM}.

When considering the case such as field trials where the temperature of each fiber segment changes independently due to the long distance, the relation between IFN and temperature becomes:
\begin{equation}
    S_{\text {fiber}}(f)=2\tau_0 f_0^2 \alpha^2 L_{\text{I}}\left(\frac{\Delta T_{\text{I}}}{\Delta t}\right)^2\cdot L f^{-2} ,
\end{equation}
which provides the former experimental observation $S_{\text {fiber}}(f)=hLf^{-2}$ in~\cite{williams2008high,newbury2007coherent} with a detailed theoretical explanation. $L_{\text{I}}$ here represents the average length of fiber whose temperature $T_{\text{I}}$ can be seen as independent to others.

%Although the function of environmental temperature with regard to time is arbitrary, it is always a continuous function. Thus the variation of the measurement clock offset with time is linear in short measurement time. 

\textbf{Delay-unsuppressed noise}. Second, a long haul of fiber means significant transfer time, so the noises generated by laser sources, links, clocks, etc. cannot be fully compensated because they could never be obtained simultaneously due to the limited speed of information transmission. The remained noise thereby is called delay-unsuppressed noise (DUN) and has been extensively observed and discussed in many transfer systems~\cite{yu2024microwave,schioppo2022comparing,droste2013optical,williams2008high,newbury2007coherent} since it casts an inevitable limit on the system performance. 

In our BAND model framework, this effect is equivalent to adding a large non-reciprocal attack in length $\Delta L_{\text{NR}}$ to a synchronization system. By considering the generalized model of insecure clock offset measurement~\cite{SM}, we can obtain:
\begin{equation}
\Delta \tau=\Delta \tau_{\text {secure}}+\Delta \tau_{\text {attack}}+\Delta \tau_{\text {DUN}}+\Delta \tau_{\text {clock delay}}.
\end{equation}
Even if $\Delta\tau_{\text{attack}}=\Delta \tau_{\text{NR}_L}/2$ is time-invariant and can be compensated, the DUN term $\Delta\tau_{\text{DUN}}=\left[\tau_{\text{channel}}(t)-\tau_{\text{channel}}(t+\Delta\tau_{\text{NR}_L})\right]/2$ and the noise generated by clock delay will never be eliminated. The differenced formulation naturally indicates the DUN is a white phase noise inherently generated by non-reciprocal delay in two directions. We can easily obtain $S_{\text{DUN}}(f)\propto (2\pi f\cdot\Delta\tau_{\text{NR}_L})^2 S_{\text{fiber}}$ according to Fourier transform, and it is consistent with previous observations~\cite{schioppo2022comparing,williams2008high}.

\textit{Conclusion}---
In this Letter, we propose an analytical security framework, BAND model, for the identification of non-reciprocal attacks. By considering physical constraints, we successfully unified different kinds of attacks and quantified their impact to the system by attack intensity. Protection schemes can thus be developed based on our model to detect or mitigate the attack on time synchronization system. Besides, by extending the BAND model to frequency signals, one can further discuss the security of the frequency transfer system against attacks~\cite{dai2024tampering}.

Beyond the attack related discussion, the BAND model is also a powerful tool to model the optical fiber, such as providing the detailed explanations for IFN and DUN, which will have a guiding significance in optimizing the performance of fiber-optic communication and time and frequency transfer. Besides, our work further indicates the potential for synchronization system to be used as distributed sensors for large-scale effect concerning vibration and temperature change, such as earthquake~\cite{marra2022optical,marra2018ultrastable}, urban transportation~\cite{wang2025laser} and terrestrial heat detection, paving
the way for a secure large-scale integrated time, sensing and communication networks.

\textit{Acknowledgements}---This work was supported by the National Natural Science Foundation of China (62201012, 62571006).

\textit{Data availability}---The data that support the findings of
this Letter are not publicly available. The data are available
from the authors upon reasonable request.

\textit{Conflict of interest statement}---None declared. 

\onecolumngrid

\section{Supplemental Material}
\subsection{APPENDIX A: INFLUENCING FACTORS OF NON-RECIPROCAL TIME DELAY IN OPTICAL FIBER\label{app1}}
The time delay of transmission $\tau$ in the optical fiber satisfies:
\begin{equation}
\tau=\frac{n L}{c}+DL\Delta\lambda,
\end{equation}
where $n$ is the refractive index, $L$ is the fiber length, $D$ is the dispersion coefficient of fiber and $\Delta\lambda$ represents the wavelength range of the transmission signal.

The main factors that influence the time delay are fiber length and temperature. The former dominates $L$ while the latter dominates $n$. Thus, by considering these two factors respectively, the change in time delay can be written as:
\begin{equation}
    \Delta \tau =\Delta \tau_L+\Delta \tau_T.
\end{equation}

The dependence on length is quite obvious:
\begin{equation}
\Delta \tau_{L}=\frac{\mathrm{d} \tau}{\mathrm{~d} L} \Delta L=\left(\frac{n}{c}+D\Delta\lambda \right) \Delta L.
\end{equation}
We take $\Delta\lambda=0.8 \mathrm{~nm}$ considering the channel we use in our dense-wavelength division multiplexing
(DWDM) module~\cite{yu2024microwave} and $n=1.5$.  Since $D=16.5~\mathrm{ps\cdot nm^{-1}\cdot km^{-1}}$ for standard single mode fiber~\cite{gruner2005dispersion}, $D\Delta\lambda\ll n/c$. We can take the approximation:
\begin{equation}
\Delta \tau_{L}\approx\dfrac{n}{c}\Delta L.
\end{equation}

The dependence on temperature is a little more complicated. We can write this dependence in detail:
\begin{equation}
\begin{aligned}
     \Delta \tau_{T}&=\Delta \tau_{T1}+\Delta \tau_{T2}\\
     &=\frac{1}{c} \frac{\mathrm{~d}(n L)}{\mathrm{d} T} \cdot \Delta T(t)+\dfrac{\mathrm{d}(DL\Delta\lambda)}{\mathrm{d}T}\cdot \Delta T(t).
\end{aligned}
\end{equation}

We will first examine $\Delta\tau_{T1}$, which represents:
\begin{equation}
\begin{aligned}
    \Delta \tau_{T1}&=\frac{1}{c} \frac{\mathrm{~d}(n L)}{\mathrm{d} T} \cdot \Delta T(t)\\
    &=\frac{L}{c} \frac{\partial n}{\partial T} \Delta T+\frac{n}{c} \frac{\partial L}{\partial T}\Delta T.
\end{aligned}
\end{equation}
Let $\alpha_{\text {th}}= \left({\partial L}/{\partial T}\right) \cdot{1}/{L}$, one yields:
\begin{equation}
\Delta  \tau_{T1}=\frac{L}{c} \frac{\partial n}{\partial T} \Delta  T(t)+\frac{n L}{c} \alpha_{\mathrm{th}}\Delta  T(t).
\end{equation}

Now let's analyze each of the three terms on the right side. The first term indicates the extra delay time brought by the refractive index's dependence on temperature. When the central wavelength is 1550 nm, ${1}/{c}\cdot  \left({\partial n}/{\partial T}\right)=35 ~\mathrm{ps} \cdot \mathrm{km}^{-1} \cdot \mathrm{~K}^{-1}$. The second term indicates the delay time brought by thermal expansion and contraction of fiber length. In our system~\cite{yu2024microwave} $\alpha_{\text{th}}=5.5\times 10^{-7}\mathrm{K}^{-1}$, which gives ${n\alpha_{\mathrm{th}}}/{c} =2.75~ \mathrm{ps} \cdot \mathrm{km}^{-1} \cdot \mathrm{~K}^{-1}$. Therefore, the dependence on temperature can be simplified as:
\begin{equation}
\Delta \tau_{T1}=\alpha L \Delta T(t),
\end{equation}
where $\alpha \approx 37.75 \mathrm{~ps} \cdot \mathrm{km}^{-1} \cdot \mathrm{~K}^{-1}$.

For $\Delta \tau_{T2}$:
\begin{equation}
\begin{aligned}
    \Delta \tau_{T2}&=\dfrac{\mathrm{d}(DL\Delta\lambda)}{\mathrm{d}T}\cdot \Delta T(t)\\
    &=L\Delta\lambda \frac{\partial D}{\partial T} \Delta T+D \alpha_{\mathrm{th}}L\Delta  T(t).
\end{aligned}
\end{equation}
Taking the typical value~\cite{hamp2002investigation} of ${\partial D}/{\partial T}\approx-1\times 10^{-3}\mathrm{~ps\cdot nm^{-1}\cdot km^{-1}\cdot K^{-1}}$, we find that $\Delta\lambda\cdot \left({\partial D}/{\partial T}\right)=-8\times 10^{-4}\mathrm{~ps\cdot km^{-1}\cdot K^{-1}}$. On the other hand, $D \alpha_{\mathrm{th}}\ll{n\alpha_{\mathrm{th}}}/{c} $. Therefore, the time delay caused by dispersion effect can be neglected. 

Based on the derivation above, we can quantitatively evaluate the non-reciprocal attack introduced by Mallory, which can be divided into two terms:
\begin{equation}
\begin{aligned}
    \Delta \tau_{\text {NR}}&=\Delta \tau_{\text {NR}_L}+\Delta \tau_{\text {NR}_T}\\
    &=\frac{n\Delta L_{\text {NR}}}{c}+\alpha L_{A B} \Delta T_{\text {NR}}+O(\Delta L \Delta T).
\end{aligned}
\end{equation}

Since $T$ and $L$ are both continuous, the non-reciprocal attack can be represented as a continuous function $\Delta \tau_{\text {attack}}(t)={\Delta \tau_{\text {NR}}}/{2}$.

\subsection{APPENDIX B: PROOF OF PROPERTIES OF EFFECTIVE ATTACK POINTS (EAPs)\label{app2}}
Since $\Delta \tau_{\text {attack}}(t)$ is continuous with respect to $t$, which can be expressed as: for $\forall \varepsilon>0$, when $\left|t-t_0\right|<\varepsilon$, there exists $\delta>0$ such that $\left|\Delta \tau_{\text {attack}}(t)-\Delta \tau_{\text {attack}}\left(t_0\right)\right|<\delta$. This is equal to the assertion: for $\forall \delta>0$, when $\left|\Delta \tau_{\text {attack}}(t)-\Delta \tau_{\text {attack}}\left(t_0\right)\right|\geq\delta$, there exists $\varepsilon>0$ such that $\left|t-t_0\right|\geq\varepsilon$.

\begin{figure}[h]
	\centering
	\includegraphics[width=0.8\linewidth]{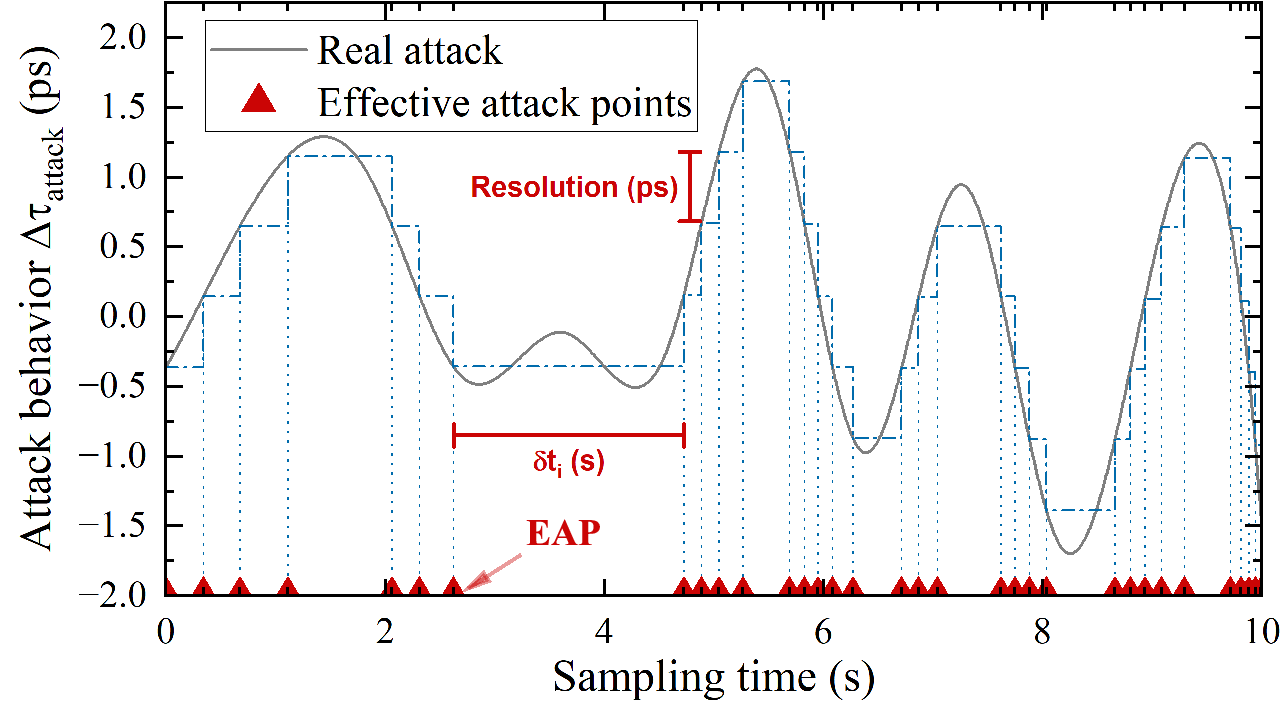}
	\caption{Visualization of EAPs. EAPs originate from discretization of real attack due to limited measurement resolution.
	}\label{EAP}
\end{figure}

This indicates that the effective attack can only be observed by user in discrete moments when resolution of measurement is not infinite. Fig.~\ref{EAP} shows the relation between real attack and effective attack points.

Now we want to prove some properties of EAP. Suppose there are an average of $\lambda$ EAPs per unit time, and the time interval between two adjacent EAPs follows an independent exponential distribution. Let the moment of the $i^{\text{th}}$ valid attack point be $t_i$, then the probability that there are $x$ valid attack points within $t$ time is:
\begin{equation}
    P(N(t)=x)=\int_0^{t_1} \int_{t_1}^{t_2} \ldots \int_{t_{x-1}}^{t}\left(e^{-\lambda t_1} \cdot \lambda e^{-\lambda\left(t_2-t_1\right)} \ldots \cdot e^{-\lambda\left(t-t_x\right)}\right) \mathrm{d} t_x \ldots \mathrm{~d} t_2 \mathrm{~d} t_1 .
\end{equation}

Since 
\begin{equation}
e^{-\lambda t_1} \cdot \lambda e^{-\lambda\left(t_2-t_1\right)} \ldots \cdot \lambda e^{-\lambda\left(t_x-t_{x-1}\right)} \cdot e^{-\lambda\left(t-t_x\right)}=\lambda^x e^{-\lambda t},
\end{equation}
\begin{equation}
\int_{t_{x-1}}^t \Delta t_x=t-t_{x-1}.
\end{equation}
Then one yields:
\begin{equation}
P(N(t)=x)=\lambda^x e^{-\lambda t} \cdot \frac{t^x}{x!}=e^{-\lambda t} \frac{(\lambda t)^x}{x!}.
\end{equation}

Therefore, the number of EAPs per unit time follows a Poisson distribution whose parameter equals $\lambda$.

\subsection{APPENDIX C: DERIVATION OF NOISE FEATURE\label{app3}}
For the attack generated by type I kernel $\Delta \tau_{\text {attack}}(t)=\sum_{k=1}^N \mathcal{T}_0 \delta\left(t-t_k\right)$,
\begin{equation}
\begin{aligned}
R_{x x}(\tau)&=\lim _{T \rightarrow \infty} \frac{\mathcal{T}_0^2}{T} \int_{-\frac{T}{2}}^{\frac{T}{2}} \Delta \tau_{\text {attack}}(t) \Delta \tau_{\text {attack}}(t+\tau) d t \\
& =\lim _{T \rightarrow \infty} \frac{\mathcal{T}_0^2}{T} \int_{-\frac{T}{2}}^{\frac{T}{2}} \sum_{k=1}^N \delta\left(t-t_k\right) \sum_{k^{\prime}=1}^N \delta\left(t-t_{k^{\prime}}+\tau\right) d t \\
& =\lim _{T \rightarrow \infty} \frac{\mathcal{T}_0^2}{T} \sum_{k^{\prime}=1}^N \sum_{k^{\prime}=1}^N \delta\left(t_k-t_{k^{\prime}}+\tau\right)\\
&=\lim _{T \rightarrow \infty} \frac{\mathcal{T}_0^2}{T} N(t) \delta(\tau).
\end{aligned}
\end{equation}

Since $N=\lambda T$, $R_{x x}(\tau)=\lambda \mathcal{T}_0^2 \delta(\tau)$, the PSD can be written as:
\begin{equation}
\begin{gathered}
S_{\text {attack}}^{2-\text {sided}}(f)=\mathcal{F}\left\{R_{x x}(\tau)\right\}=\lambda \mathcal{T}_0{ }^2, \\
S_{\text {attack}}(f)=2 S_{\text {attack}}^{2-\text {sided}}=2 \lambda \mathcal{T}_0^2,
\end{gathered}
\end{equation}
which corresponds to a white phase noise.

On the other hand, the Fourier transform satisfies:
\begin{equation}
\begin{gathered}
\mathcal{F}\left\{\frac{d x(t)}{d t}\right\}=-\mathrm{i} 2 \pi f \cdot \mathcal{F}\{x(t)\}, \\
S(f)=\lim _{T \rightarrow \infty} \frac{1}{T}|\mathcal{F}\{x(t)\}|^2.
\end{gathered}
\end{equation}

Therefore, the PSD of attack generated by type II kernel $\Delta \tau_{\text {attack}}(t)=\sum_{k=1}^N \mathcal{T}_0 U\left(t-t_k\right)$ is:
\begin{equation}
S(f)=\lim _{T \rightarrow \infty} \frac{1}{T}|\mathcal{F}\{x(t)\}|^2=\frac{2 \lambda \mathcal{T}_0^2}{|-2 i \pi f|^2}=\frac{\lambda \mathcal{T}_0^2}{2 \pi^2 f^2},
\end{equation}
which corresponds to a white frequency noise.

\subsection{APPENDIX D: CHOICE OF TDEV FITTING RANGE\label{app5}}
Tab.~\ref{tab:noise_parameters} (derived based on~\cite{calosso2016avoiding}) gives the relation between PSD of power-law noise $S_{\varphi}(f)$ and the corresponding TDEV coefficient $C_{\text{noise}}$. We will use the table to illustrate our selection principles of the fitting range.

\begin{table}[htbp]
\caption{\label{tab:noise_parameters}
Formula-based correspondence between PSD and TDEV coefficient of power-law noise.
}
\begin{ruledtabular}
\begin{tabular}{lccccc}
Noise Type & White PN & Flicker PN & White FN & Flicker FN & Random Walk FN \\
\colrule
$\boldsymbol{S}_{\boldsymbol{\varphi}}(\boldsymbol{f})$ & 
$b_0$ & 
$b_{-1} f^{-1}$ & 
$b_{-2} f^{-2}$ & 
$b_{-3} f^{-3}$ & 
$b_{-4} f^{-4}$ \\
[1.5ex] % 增加一点行间距，防止平方根符号与上一行重叠
$\boldsymbol{C}_{\text {noise }}$ & 
$\sqrt{\dfrac{1}{8 \pi^2} \dfrac{b_0}{f_0^2}} \tau^{-\frac{1}{2}}$ & 
$\sqrt{\dfrac{\ln (256 / 27)}{8 \pi^2} \dfrac{b_{-1}}{f_0^2}}$ & 
$\sqrt{\dfrac{1}{12} \dfrac{b_{-2}}{f_0^2}} \tau^{\frac{1}{2}}$ & 
$\sqrt{\dfrac{9}{20} \ln (2) \dfrac{b_{-3}}{f_0^2} \tau}$ & 
$\sqrt{\dfrac{11 \pi^2}{60} \dfrac{b_{-4}}{f_0^2}} \tau^{\frac{3}{2}}$ \\
\end{tabular}
\end{ruledtabular}
\end{table}

Since the white phase noise is the only power-law noise whose TDEV is negatively correlated with $\mathrm{log}(\tau)$ among the five noises, the extraction of it should use the part with a smaller $\tau$. In our first experiment, with a sampling rate of $1$ kHz, we used TDEV from $0.001$ s to $1$ s to calibrate the intensity of white phase noise.

\begin{figure}[h]
	\centering
	\includegraphics[width= 0.8\linewidth]{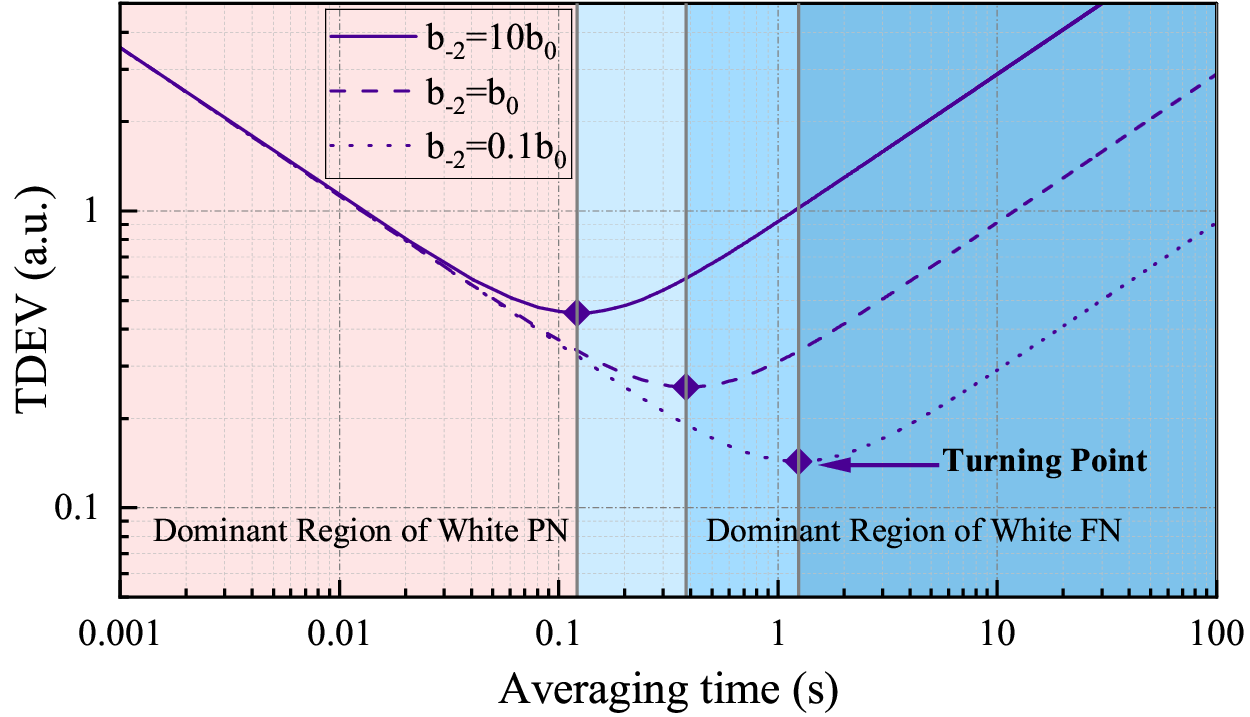}
	\caption{Dominant regions of white phase noise and white frequency noise are divided by the turning point, which depends on the ratio of intensity of these two noise.
	}\label{fitting range}
\end{figure}

For white frequency noise, its dominant region of TDEV is determined by the intensity ratio of it to the white phase noise. As shown in Fig.~\ref{fitting range}, the turning point $\tau_{\text{turn}}={C_{-2}}/{C_{0}}$ divides the two dominant regions. The greater the white frequency noise relative to the white phase noise, the smaller the $\tau_{\text{turn}}$. If one wants to extract smaller white frequency noise, a region with greater $\tau$ is required. In this paper, it is planned to extract the white frequency noise, whose intensity is approximately between 0.1 and 10 times of the white phase noise of the secure system. Therefore, TDEV ranging from 0.1 s to 10 s is selected for its fitting. In the temperature drift scenario, we choose 1 s to 100 s for fitting, corresponding to a smaller white frequency noise. However, when it gets smaller, the dominant region may overlap with higher power-law noise, and thus the extraction will be more difficult. Therefore, the noise structure of time synchronization systems determines their security, and reasonable selection of the fitting range is crucial to the accurate extraction of the attack.

\subsection{APPENDIX E: GENERALIZED MODEL OF INSECURE CLOCK OFFSET MEASUREMENT\label{app4}}
Let the local time of Alice to be
\begin{equation}
\begin{aligned}
   T_A(t)&=\frac{\phi_{A 0}}{2 \pi f_0}+\frac{1}{2 \pi f_0} \int_{t_0}^t\left(f_0+\Delta f_B(t)\right) \mathrm{d} t\\
    &=T_A\left(t_0\right)+\frac{1}{2 \pi f_0} \int_{t_0}^t\left(f_0+\Delta f_A(t)\right) \mathrm{d} t
\end{aligned}
\end{equation}
with a clock frequency jitter $\Delta f_A(t)$. And the local time of Bob is
\begin{equation}
T_B(t)=T_B\left(t_0\right)+\frac{1}{2 \pi f_0} \int_{t_0}^t\left(f_0+\Delta f_B(t)\right) \mathrm{d} t
\end{equation}
with a clock frequency jitter $\Delta f_B(t)$.

If Alice and Bob want to synchronize their time at $t_0$, then when the system is secure, two time differences acquired according to Einstein protocol are
\begin{equation}
\begin{aligned}
& \Delta T_{A \rightarrow B}\left(t_0\right)=T_A\left(t_0\right)+\tau_{\text {channel}}\left(t_0\right)-\left[T_B\left(t_0\right)+\frac{1}{f_0} \int_{t_0}^{t_0+\tau_{\text {channel}}}\left(f_0+\Delta f_B(t)\right) \mathrm{d} t\right], \\
& \Delta T_{B \rightarrow A}\left(t_0\right)=T_B\left(t_0\right)+\tau_{\text {channel}}\left(t_0\right)-\left[T_A\left(t_0\right)+\frac{1}{f_0} \int_{t_0}^{t_0+\tau_{\text {channel}}}\left(f_0+\Delta f_A(t)\right) \mathrm{d} t\right].
\end{aligned}
\end{equation}

After subtraction, they can obtain the result:
\begin{equation}
\Delta \tau_{\text {secure}}=\frac{\Delta T_{A \rightarrow B}-\Delta T_{B \rightarrow A}}{2}=\left(T_A\left(t_0\right)-T_B\left(t_0\right)\right)+\frac{1}{2 f_0} \int_{t_0}^{t_0+\tau_{\text {channel}}}\left(\Delta f_A(t)-\Delta f_B(t)\right) \mathrm{d} t
,\end{equation}
where $\left(T_A\left(t_0\right)-T_B\left(t_0\right)\right)$ is the clock offset Alice and Bob need, and $\frac{1}{2 f_0} \int_{t_0}^{t_0+\tau_{\text {channel}}}\left(\Delta f_A(t)-\Delta f_B(t)\right) \mathrm{d} t$ is the noise when the system is secure.

Now suppose the system is under attack. If Alice and Bob continue to synchronize their time, then the two time differences they measure become: 
\begin{equation}
\Delta T_{A \rightarrow B}\left(t_0\right)=T_A\left(t_0\right)+\tau_{\text {channel}}\left(t_0\right)+\Delta \tau_{\text {NR}}\left(t_0\right)-\left[T_B\left(t_0\right)+\frac{1}{f_0} \int_{t_0}^{t_0+\tau_{\text {channel}}+\Delta \tau_{\text {NR}}}\left(f_0+\Delta f_B(t)\right) \mathrm{d} t\right],
\end{equation}
\begin{equation}
\Delta T_{B \rightarrow A}\left(t_0\right)=T_B\left(t_0\right)+\tau_{\text {channel}}\left(t_0\right)-\left[T_A\left(t_0\right)+\frac{1}{f_0} \int_{t_0}^{t_0+\tau_{\text {channel}}}\left(f_0+\Delta f_B(t)\right) \mathrm{d} t\right].
\end{equation}

After subtraction, one can obtain:
\begin{equation}
\begin{aligned}
\Delta \tau&=\frac{\Delta T_{A \rightarrow B}\left(t_0\right)-\Delta T_{B \rightarrow A}\left(t_0+\Delta \tau_{\text {NR}}\right)}{2}\\
& =\left[\left(T_A\left(t_0\right)-T_B\left(t_0\right)\right)+\frac{1}{2 f_0} \int_{t_0}^{t_0+\tau_{\text {channel}}}\left(\Delta f_A(t)-\Delta f_B(t)\right) \mathrm{d} t\right] \\
& +\frac{\Delta \tau_{\text {NR}}}{2}+\frac{\left[\tau_{\text {channel}}\left(t_0\right)-\tau_{\text {channel}}\left(t_0+\Delta \tau_{\text {NR}}\right)\right]}{2} \\
& +\left[\frac{1}{2 f_0} \int_{t_0+\tau_{\text {channel}}}^{t_0+\tau_{\text {channel}}+\Delta \tau_{\text {NR}}}\left(\Delta f_A(t)-\Delta f_B(t)\right) \mathrm{d} t-\frac{1}{2 f_0} \int_{t_0}^{t_0+\Delta \tau_{\text {NR}}} \Delta f_B(t) \mathrm{d} t\right] \\
& =\Delta \tau_{\text {secure}}+\Delta \tau_{\text {attack}}+\Delta \tau_{\text {DUN}}+\Delta \tau_{\text {clock delay}}.
\end{aligned}
\end{equation}
where the first two terms are the same as in secure case. The attack introduces three more terms: ${\Delta \tau_{\text {NR}}}/{2}$ represents the deviation and noise brought by attack itself; ${\left[\tau_{\text {channel}}\left(t_0\right)-\tau_{\text {channel}}\left(t_0+\Delta \tau_{\text{NR}}\right)\right]}/{2}$ represents the delay-unsuppressed noise (DUN) introduced by non-reciprocal channel; the last term is the clock delay noise due to different arrival time of light pulses.

When the delay is small, DUN noise and clock delay noise can be dismissed. Then $\Delta\tau$ converges to:
\begin{equation}
\begin{aligned}
    \Delta \tau &\approx\left[\left(T_A\left(t_0\right)-T_B\left(t_0\right)\right)+\frac{1}{2 f_0} \int_{t_0}^{t_0+\tau_{\text {channel}}}\left(\Delta f_A(t)-\Delta f_B(t)\right) \mathrm{d} t\right]+\frac{\Delta \tau_{\text {NR}}}{2}\\
    &=\Delta \tau_{\text {secure}}+\Delta \tau_{\text {attack}}.
\end{aligned}
\end{equation}

\subsection{APPENDIX F: CALCULATION OF THEORETICAL ATTACK INTENSITY\label{app6}}
Attack intensity under the constraint of limited resolution is written as $I=\lambda \mathcal{T}^2_0$. However, in a real synchronization system, due to the limited sampling rate $s$, EAPs become a subset of sampling points and $\Delta \tau_{\text {attack}}$ will be discretized by factor $\mathcal{T}_i$, which is greater than the resolution $\mathcal{T}_0$. Therefore, the theoretical attack intensity should be $I_{\text{theo}}=\lambda\mathbb{E}\left\{\mathcal{T}^2\right\}$.

For the first experiment where Mallory applies disturbance to the system, the attack kernel is deviated from an ideal type I kernel by a factor $1/s$. The theoretical attack intensity is therefore $I_{theo}=\lambda\mathbb{E}\left\{\mathcal{T}^2\right\}/s^2$. The comparison between theoretical and experimental data is shown in Tab.~\ref{tab:k1data}.
\begin{table}[h]
\caption{\label{tab:k1data}
The experiment data of instantaneous disturbance attack.
}
\begin{ruledtabular}
\begin{tabular}{lccc}
& \multicolumn{1}{c}{$\boldsymbol{I}_{\text {theo}}(\boldsymbol{s})$} &
\multicolumn{1}{c}{$\boldsymbol{I}_{\text {detect}}(\boldsymbol{s})$} &
\multicolumn{1}{c}{$\boldsymbol{\eta}(\%)$} \\
\colrule
Background & $\mathrm{N} / \mathrm{A}$ & $6.9607 \mathrm{e}-28$ & $\mathrm{~N} / \mathrm{A}$ \\
Experiment & $3.7495 \mathrm{e}-28$ & $1.0891 \mathrm{e}-27$ & 4.822 \\
\end{tabular}
\end{ruledtabular}
\end{table}

For the second experiment where Mallory uses optical delay line, the theoretical attack intensity can be calculated directly from $I_{\text{theo}}=\lambda\mathbb{E}\left\{\mathcal{T}^2\right\}$. The comparison between theoretical and experimental data is shown in Tab.~\ref{tab:k2data}.

\begin{table}[h]
\caption{\label{tab:k2data}
The experiment data of accumulating delay attack.
}
\begin{ruledtabular}
\begin{tabular}{lccc}
& \multicolumn{1}{c}{$\boldsymbol{I}_{\text {theo}}(\boldsymbol{s})$} & 
\multicolumn{1}{c}{$\boldsymbol{I}_{\text {detect}}(\boldsymbol{s})$} & 
\multicolumn{1}{c}{$\boldsymbol{\eta}(\%)$} \\
\colrule
Background & $\mathrm{N} / \mathrm{A}$ & 0 & $\mathrm{~N} / \mathrm{A}$ \\
Experiment & $1.1578 \mathrm{e}-25$ & $1.2136 \mathrm{e}-25$ & 4.819 \\
\end{tabular}
\end{ruledtabular}
\end{table}

For the third experiment where Mallory adds a long section of fiber, the fluctuating temperature will have a continuous influence on the system.  Thus $\lambda$ equals the sampling rate $s$ and the attack is generated by type II kernel with amplitude $\left\{t_{i+1}-t_i\right\}$. Let $\left\{y_i \left\lvert\, y_i=\frac{t_{i+1}-t_i}{\tau_0}\right.\right\}$, following the derivation in Appendix C, the theoretical attack intensity can be derived as: 
\begin{equation}
    I_{\text{theo}}=\lambda \mathbb{E}\left\{\mathcal{T}^2\right\}=\dfrac{1}{\tau_0} \cdot \sum_{i=1}^{N-1} \frac{\left(t_{i+1}-t_{i}\right)^2}{(N-1)}=\tau_0\cdot\mathbb{E}\{y^2\}.\label{eq:itheo_of_temp}
\end{equation}
This result is natural and is the so-called intrinsic fiber noise (IFN)~\cite{williams2008high,newbury2007coherent}.

%In real system, one need to exclude the noise when system is secure, that is to say:
%\begin{equation}
%I_{\text {theo }}= \lambda \mathbb{E}\left\{\mathcal{T}^2\right\}_{\text{attack}}-\lambda \mathbb{E}\left\{\mathcal{T}^2\right\}_{\text{secure}}.
%\end{equation}

%After considering the non-homogeneous correction, the theoretical attack intensity for 2nd-order drift can be derived as:
%\begin{equation}
%\begin{gathered}
%\lambda \mathbb{E}\left\{\mathcal{T}^2\right\}=\frac{1}{\tau_0} \cdot \sum_{i=1}^{N-2} \frac{\left[\left(t_{i+2}-t_{i+1}\right)-\left(t_{i+1}-t_i\right)\right]^2}{(N-2)}=\frac{\tau_0^2}{\tau_0} \cdot \sum_{i=1}^{N-2} \frac{\left[\left(\phi_{i+2}-\phi_{i+1}\right)-\left(\phi_{i+1}-\phi_i\right)\right]^2}{\left(\omega_0 \tau_0\right)^2(N-2)} \\
%=2 \tau_0 \cdot \sum_{i=1}^{N-2} \frac{\left(y_{i+1}-y_i\right)^2}{2(N-2)}=2 \tau_0 \cdot \text { AVAR } @ \tau_0
%\end{gathered}
%\end{equation}
%where the difference of  $y$ is used to eliminate the drift trend, since in this case the frequency is no longer stationary. 

The comparison between theoretical and experimental data is shown in Tab.~\ref{tab:tempdata}. Note that non-linear drift violates our postulate 3, and thus needs a non-homogeneous correction. The correction in this scenario extracts the stationary component of non-linear temperature drift by eliminating the trend of frequency, which can be written as:
\begin{equation}
    I_{\text{theo}}^{\prime}=\lambda \mathbb{E}\left\{\mathcal{T_{\text{stationary}}}^2\right\}=\dfrac{\tau_0}{2} \cdot \sum_{i=1}^{N-2} \frac{\left(y_{i+1}-y_{i}\right)^2}{(N-2)},
\end{equation}
where the factor $2$ in the denominator is added since $I_{\text{theo}}^{\prime}$ should converge to Eq.~\eqref{eq:itheo_of_temp} when $\left\{y_i\right\}$ is not correlated to $\left\{y_{i+1}\right\}$.

\begin{table}[htbp]
\caption{\label{tab:tempdata}
The experiment data of temperature drift.
}
\begin{ruledtabular}
\begin{tabular}{lccc}
& \multicolumn{1}{c}{$I_{\text {theo}}(s)$} & 
\multicolumn{1}{c}{$I_{\text {detect}}(s)$} & 
\multicolumn{1}{c}{$\boldsymbol{\eta}$ (\%)} \\
\colrule
Background & N/A & 0 & N/A \\
Linear fast drift & $2.71 \mathrm{e}-26$ & $2.62 \mathrm{e}-26$ & 3.18 \\
Linear slow drift & $2.86 \mathrm{e}-26$ & $3.07 \mathrm{e}-26$ & 7.34 \\
Non-linear drift (before correction) & $6.33 \mathrm{e}-26$ & $1.73 \mathrm{e}-26$ & 72.67 \\
Non-linear drift (after correction) & $1.90 \mathrm{e}-26$ & $1.73 \mathrm{e}-26$ & 9.19 \\
\end{tabular}
\end{ruledtabular}
\end{table}

%Since the noise is white frequency noise, then: 
%\begin{equation}
%\sum_{i=1}^{N-2} \frac{\left(y_{i+1}-y_i\right)^2}{2(N-2)}=\mathbb{E}\left\{y^2\right\}-\sum_{i=1}^{N-2} \frac{y_{i+1} y_i}{(N-2)}=\mathbb{E}\left\{y^2\right\}-\operatorname{Corr}\left(\tau_0\right)=\mathbb{E}\left\{y^2\right\}
%\end{equation}

When the temperature of every section of fiber is aligned, the PSD of IFN can be written as:
\begin{equation}
\begin{aligned}
S_{\text {fiber}}(f)=2 \tau_0 \cdot\left(\frac{f_0}{f}\right)^2 \cdot \mathbb{E}\left\{y^2\right\}=2 \tau_0\left(\frac{f_0}{f}\right)^2 \alpha^2 L^2 \mathbb{E}\left\{\left(\frac{\Delta T}{\Delta t}\right)^2\right\}\propto L^2 \mathbb{E}\left\{\left(\frac{\Delta T}{\Delta t}\right)^2\right\},
\end{aligned}
\end{equation}
which shows clearly how fiber length and temperature fluctuation contribute to the noise.

However, in many realistic cases, the fiber is laid over a long distance, thus the temperature of each section of fiber changes independently. In these cases,  $y$ can be seen as a random walk, therefore the total contribution to the measured $y$ is proportional to fiber length, that is to say: 
\begin{equation}
\mathbb{E}\left\{y^2\right\}=y_I^2 \mathbb{E}\left\{\left(\frac{y}{y_I}\right)^2\right\}=y_I^2 \mathbb{E}\left\{\left(\frac{L \Delta T}{L_I \Delta T_I}\right)^2\right\}\approx y_I^2 \cdot\frac{L}{L_I}=\alpha^2 L L_I \left(\frac{\Delta T_I}{\Delta t}\right)^2,
\end{equation}
where $L_I$ is the characteristic thermal length of fiber that has independent response to the environment temperature $T_I$. This is quite different from the former cases when the temperature is aligned since the noise is proportional to $L$ rather than $L^2$. The observations of previous works~\cite{williams2008high,newbury2007coherent,schioppo2022comparing,yu2024microwave} using outfield fiber link support our theory.

\bibliographystyle{apsrev4-2}
\bibliography{nsr_sample}

%\begin{thebibliography}{99}  
%
%
%\end{thebibliography}
%
%

\end{document}